\documentclass[%
 reprint,
superscriptaddress,
 amsmath,amssymb,
 aps,
]{revtex4-2}

\usepackage{graphicx}%
\usepackage{dcolumn}%
\usepackage{bm}%

\usepackage{amsthm}
\usepackage{appendix}
\usepackage{graphicx}
\usepackage{algorithmic}
\usepackage{algorithm}
\usepackage[roman]{complexity}
\usepackage{mathrsfs}
\usepackage{braket}
\usepackage{xcolor}
\usepackage{tikz}
\usetikzlibrary{quantikz}

\usepackage{hyperref}
\usepackage[capitalize]{cleveref}

\newtheorem{thm}{\protect\theoremname}
\theoremstyle{plain}
\newtheorem{lem}[thm]{\protect\lemmaname}
\theoremstyle{plain}

\theoremstyle{plain}
\newtheorem*{lem*}{\protect\lemmaname}
\theoremstyle{plain}

\theoremstyle{plain}
\newtheorem{cor}[thm]{\protect\corollaryname}
\newtheorem{assump}{Condition}
\newtheorem{defn}[thm]{Definition}

\newcommand{\norm}[1]{\left\lVert#1\right\rVert}

\usepackage[USenglish]{babel}
  \providecommand{\corollaryname}{Corollary}
  \providecommand{\lemmaname}{Lemma}
  \providecommand{\propositionname}{Proposition}
  \providecommand{\remarkname}{Remark}
\providecommand{\theoremname}{Theorem}

\usepackage[normalem]{ulem}

\newcommand{\Or}{\mathcal{O}}
\newcommand{\RR}{\mathbb{R}}

\newcommand{\tr}{\mathrm{tr}}

\renewcommand{\Re}{\operatorname{Re}}
\renewcommand{\Im}{\operatorname{Im}}

\usepackage{todonotes}

\newcommand{\REV}[1]{#1}
\newcommand{\REVsec}[1]{#1}

\newcommand{\ud}{\,\mathrm{d}}

\IfFileExists{mathabx.sty}%
  {\DeclareFontFamily{U}{mathx}{\hyphenchar\font45}%
   \DeclareFontShape{U}{mathx}{m}{n}{<->mathx10}{}%
   \DeclareSymbolFont{mathx}{U}{mathx}{m}{n}%
   \DeclareFontSubstitution{U}{mathx}{m}{n}%
   \DeclareMathAccent{\widebar}{0}{mathx}{"73}%
}{%
  \PackageWarning{mathabx}{%
    Package mathabx not available, therefore\MessageBreak substituting
    widebar with overline\MessageBreak }%
  \newcommand{\widebar}[1]{\overline{#1}}%
}

\newcommand{\mc}[1]{\mathcal{#1}}

\newcommand{\mf}[1]{\mathsf{#1}}

\newcommand{\veps}{\varepsilon}

\newcommand{\average}[1]{\langle#1\rangle}
\newcommand{\Average}[1]{\left\langle#1\right\rangle}

\renewcommand{\Re}{\mathrm{Re}} 
\renewcommand{\Im}{\mathrm{Im}}

\begin{document}

\title{Mixing Time of Open Quantum Systems via Hypocoercivity}

\author{Di Fang}
\affiliation{Department of Mathematics, Duke University}
\affiliation{Duke Quantum Center, Duke University}

\author{Jianfeng Lu}
\affiliation{Department of Mathematics, Duke University}
\affiliation{Department of Physics, Duke University}
\affiliation{Department of Chemistry, Duke University}

\author{Yu Tong}
\affiliation{Institute for Quantum Information and Matter, California Institute of Technology}
\affiliation{Department of Mathematics, Duke University}
\affiliation{Department of Electrical and Computer Engineering, Duke University}

\date{\today}

\begin{abstract}

Understanding the mixing of open quantum systems is a fundamental problem in physics and quantum information science. Existing approaches for estimating the mixing time often rely on the spectral gap estimation of the Lindbladian generator, which can be challenging to obtain in practice. We propose a novel theoretical framework to estimate the mixing time of open quantum systems that treats the Hamiltonian and dissipative part separately, thus circumventing the need for a priori estimation of the spectral gap of the full Lindbladian generator. \REVsec{This framework yields mixing time estimates for a class of quantum systems that are otherwise hard to analyze, even though it does not apply to arbitrary Lindbladians.} The technique is based on the construction of an energy functional inspired by the hypocoercivity of (classical) kinetic theory.
\end{abstract}

\maketitle

The Lindblad equation describes the dynamics of open quantum systems under Markovianity~\cite{Lindblad1976,GoriniKossakowskiSudarshan1976}, providing a useful tool for understanding the dynamics of quantum states in the presence of weak environmental interactions. A wide range of physical systems that involve such environmental interactions, from quantum optics to condensed matter physics, can be described by the Lindblad equation. For a quantum system with Hamiltonian $H$ that is coupled to an environment through jump operators $V_j$, the Lindbladian super-operator, in the Heisenberg picture, takes the following form: 
\begin{equation}
\label{eq:lindblad_heisenberg}
    \mc{L} A = \underbrace{i [ H, A]}_{ =: \mc{H}} + \underbrace{\sum_j V_j^{\dag} [A, V_j] + [V_j^{\dag}, A] V_j}_{=:\mc{D}}.
\end{equation}
The evolution of an observable $A$ under the Heisenberg picture is described by the Lindblad equation $
\frac{\ud}{\ud t}A = \mc{L} A.
$
We can equivalently consider the evolution of a quantum state in the Schr\"{o}dinger picture, for which we will need to use the dual of the Lindbladian
\begin{equation}
\label{eq:lindblad_Schrodinger}
    \mc{L}^\star \rho = -i [ H, \rho] + \sum_j [V_j\rho, V_j^{\dag}] + [V_j, \rho V_j^\dag].
\end{equation}
Then the evolution of a quantum state in the Schr\"{o}dinger picture is described by
$
\frac{\ud}{\ud t}\rho = \mc{L}^\star \rho.
$
We will decompose the Lindbladian $\mc{L}$ into two parts: $\mc{L}=\mc{H}+\mc{D}$.
The first part $\mc{H}$, which we will call the \emph{Hamiltonian part}, describes the unitary dynamics of the quantum system in isolation, while the second part $\mc{D}$ describes dissipation as a result of interaction with the environment, and we will call it the \emph{dissipative part}.

The significance of the Lindblad equation extends beyond its ability to describe physical reality. Carefully designed dissipation can be used to achieve universal quantum computation \cite{VerstraeteWolfCirac2009}, and is useful in quantum field theories simulation~\cite{OsborneEisertVerstraete2010,VerstraeteCirac2010}.
Recent years have witnessed significant advances in employing the Lindblad equation to develop algorithms to prepare and sample from the thermal states~\cite{ChenKastoryanoGilyen2023,ChenKastoryanoBrandaoGilyen2023,RallWangWocjan2023,DingLiLin2024}, to prepare the ground state~\cite{DingChenLin2023}, to find local minima in quantum systems~\cite{ChenHuangPreskillZhou2023}, and even for scientific computing purposes beyond the quantum origin, such as optimization problems~\cite{ChenLuWangLiuLi2023}. 
These algorithms recast the problems of interest into tasks of preparing the stationary state of a Lindblad equation. This can then be efficiently addressed using state-of-the-art Lindbladian solvers, such as those developed in~\cite{CleveWang2017, WocjanTemme2023, ChenKastoryanoBrandaoGilyen2023, ChenKastoryanoGilyen2023, LiWang2023, DingLiLin2023,PocrnicSegalWiebe2024}.

At the center of the studies of Lindbladian dynamics is the mixing time, the time it takes for the physical system to reach to its equilibrium.
An estimation of the mixing time typically plays a crucial role in many of these algorithmic applications to yield efficient quantum algorithms. For example, the mixing time may appear as a parameter in the final cost estimate for these algorithms: The algorithm is efficient provided that the mixing time scales polynomially with the size of the system.

Significant progress on the analysis and understanding of the mixing time has been made over the past years. One typically starts with a spectral gap estimation of the Lindbladian operator $\mathcal{L}$. 
For certain special cases, this gap can be directly estimated \cite{temme2013lower, KastoryanoBrandao2016,BarthelZhang2022superoperator}, 
which then gives us information about the mixing time.
Certain physically motivated assumptions, such as the eigenstate thermalization hypothesis (ETH), can also be used to provide gap estimates \cite{ChenBrandao2021}.
Moreover, when there is a finite spectral gap of $\mathcal{L}$, the state-of-the-art developments of the modified log-Sobolev inequality~\cite{KastoryanoTemme2013,CapelRouzeFranca2020, BardetCapelGaoEtAl2023, BardetCapelLuciaEtAl2021,LiRouze2022,LiJungeLaRacuenteLi2022, LiLu2023,BardetCambyse2022} 
provide means to tighten the bound even further by providing a $\polylog(N)$ mixing time (i.e., rapid mixing) estimate for some special classes of Hamiltonians, such as the 1D commuting Hamiltonians of system size $N$. 
Recently, the spectral gap of certain modified Davies generators has been associated with the spectral gap of a quasi-local frustration-free parent Hamiltonian \cite{ChenKastoryanoGilyen2023}, leading to mixing time estimation via analyzing the parent Hamiltonian spectral gap \cite{RouzeFrancaAlhambra2024}. These works generally focus on preparing the Gibbs state through the Lindbladian dynamics, but the Gibbs state can also be prepared using approximate Markovianity \cite{BrandaoKastoryano2019}, utilizing the lack of entanglement at a sufficiently high temperature \cite{BakshiLiuMoitraTang2024}, through a quantum Metropolis algorithm analogous to a discrete-time random walk \cite{TemmeOsborneVollbrechtPoulinVerstraete2011}, or through matrix function and post-selection \cite{ChowdhurySomma2017,HolmesMuraleedharanEtAl2022,PoulinWocjan2009,ChowdhuryLowWiebe2020variational,PoulinWocjan2009,ChowdhuryLowWiebe2020variational,TongAnWiebeLin2021fast}.
  
Despite the recent progress, obtaining estimates of the spectral gap of the Lindbladian operator is typically difficult and can sometimes be the bottleneck of the analysis. Note that the Lindbladian operator consists of a Hamiltonian part $\mathcal{H}$ and a dissipation part $\mathcal{D}$. It is often easier to deal with $\mathcal{H}$ and $\mathcal{D}$ separately: In many cases, the dissipative part $\mathcal{D}$ can be easier to analyze, and sometimes we can even obtain the entire spectrum. Examples include when $\mc{D}$ describes Pauli noise. Another merit to consider $\mathcal{D}$ separately is that it may only act on part of the system, such as a few qubits, which could facilitate efficient algorithmic estimation.
However, the dissipative part alone is often insufficient to guarantee the mixing of the entire dynamics. In fact, one can easily find examples where the entire dynamics converges to a unique fixed point while the dissipative part (such as Pauli noise) does not. This raises the natural question:
\begin{center}
\textit{
Can we use information about the dissipative part $\mathcal{D}$ and the Hamiltonian part $\mathcal{H}$ separately to yield a mixing time estimation?
}\end{center} 
In particular, we would like to investigate if there are conditions imposed on $\mathcal{H}$ and $\mathcal{D}$ 
such that we can avoid performing a full gap estimation of $\mathcal{L}$ yet still be able to describe the convergence of the dynamics quantitatively.

Our work answers this question affirmatively. We provide a theoretical framework and propose a set of conditions on the operators $\mathcal{H}$ and $\mathcal{D}$ such that we can still establish the exponential convergence of the Lindbladian dynamics to its equilibrium \REVsec{for a class of Lindbladians}. Moreover, we provide a quantitative estimate of the convergence rate (Theorem \ref{thm:main}) and hence the mixing time (\cref{coro:mixing_time_optimized}). We then provide a number of physical examples that satisfy our conditions, including the transverse field Ising model, Heisenberg model, and quantum walk, with some Pauli noise. The mathematical framework and the proof are heavily inspired by the hypocoercivity established in (classical) kinetic theory~\cite{Villani2009,Villani2006,mouhot2006quantitative,  dolbeault2009hypocoercivity, DolbeaultMouhotSchmeiser2015} that provides the global convergence of the Boltzmann equation and Fokker-Planck equations.
The main idea is to construct a Lyapunov functional $\mf{L}[A]$ that captures how the Hamiltonian part also contributes to the dissipation.

\medskip
\noindent \textit{Hypocoercivity and Physical Intuition.---}
Hypocoercivity has been a successful and celebrated concept in kinetic theory (see, e.g., the works~\cite{Villani2006,Villani2009, mouhot2006quantitative, dolbeault2009hypocoercivity, DolbeaultMouhotSchmeiser2015}). Roughly speaking, coercivity refers to system with spectral gap, while hypocoercivity aims to establish convergence without the spectral gap assumption. 
We discuss the physical intuition of hypocoercivity and illustrate the underlying main idea in this section using a simple one-qubit toy example, before going into details of the theoretical framework. 

\begin{figure}[t]
    \centering
    \includegraphics[width = 0.3\textwidth]{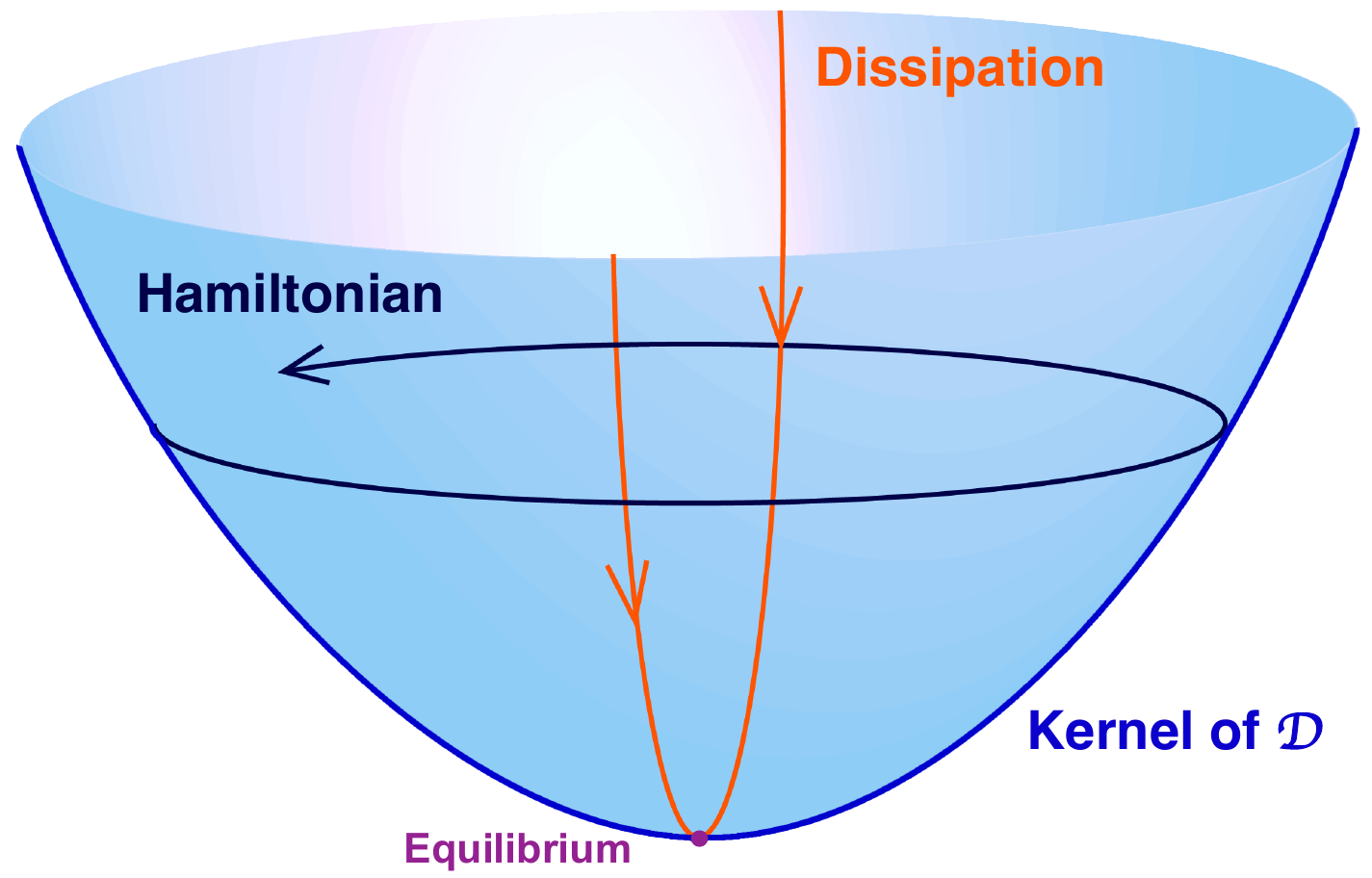}
    \caption{The intuition behind hypocoercivity is depicted as follows: The blue curve represents the kernel of \(\mathcal{D}\), that is, \(\ker\mathcal{D}\), on which $\mc{D}$ has no effect. The Hamiltonian component acts as a mixing mechanism that pushes the system out of the kernel of \(\mathcal{D}\), so that the dissipative component can damp the dynamics, guiding the system towards equilibrium.}
    \label{fig:intuition}
\end{figure}

Let us consider the Lindbladian dynamics $e^{t\mathcal{L}} = e^{t\mathcal{H} + t\mathcal{D}}$. We can adopt the Trotterization perspective, where $e^{t\mathcal{L}} = \lim_{L \to \infty}\left( e^{t\mathcal{H}/L}e^{t\mathcal{D}/L} \right)^L$. Any matrix $A$ can be decomposed into components inside and outside the kernel of $\mathcal{D}$, denoted as $A = \mathcal{P} A + (\mathcal{I-P})A$, where $\mathcal{P}$ is the projection onto the kernel of  $\mathcal{D}$. The dissipative nature of $\mathcal{D}$ ensures that the dynamics exponentially damps $(\mathcal{I-P})A$ (we will later formalize this in \cref{assump:micro}). For $\mathcal{P}A$, we observe that
$e^{t\mathcal{H}/L} \mathcal{P}A = \mathcal{P}A + \mathcal{H}\mathcal{P}At/L + \Or(t^2/L^2),$
where the second term on the right-hand side is equal to $(\mathcal{I}-\mathcal{P})\mathcal{H}\mathcal{P}At/L$ if $\mathcal{P}\mathcal{H}\mathcal{P} =0$ (later formalized in \cref{assump:C}). Consequently, the $\mathcal{P}A$ part is being driven to the image of $\mathcal{I}-\mathcal{P}$, i.e., the orthogonal complement of $\ker{\mathcal{D}}$, which will then be damped by $\mathcal{D}$ in the next step. This iterative process continues until the system resides in the kernel of both $\mathcal{H}$ and $\mathcal{D}$, representing a fixed point of $\mathcal{L}$. We graphically illustrate the intuition in \cref{fig:intuition}.

On a more concrete level, let us consider a single-qubit example with Hamiltonian and a jump operator given by
$H = X$, $V = \ket{0}\bra{0},$
where $X$ denotes the Pauli-$X$ operator.
Explicit calculation yields that $\ker \mc{D}$ consists of diagonal matrix, and for $A = \bigl(\begin{smallmatrix} a & b \\ c & d \end{smallmatrix}\bigr)$, 
$
\mc{P} A = 
\bigl(\begin{smallmatrix}
    a & 0 \\
    0 & d
\end{smallmatrix}\bigr).
$
Notice that in this example, $\mathcal{H}$ transforms a diagonal matrix into an off-diagonal one, so that $\mathcal{P}\mathcal{H}\mathcal{P} = 0$. We can also compute that
$
e^{t\mathcal{D}} A =
\bigl(\begin{smallmatrix}
    a & e^{-t} b \\
    e^{-t} c  & d
\end{smallmatrix}\bigr),
$
where everything outside of $\ker\mc{D}$ is damped exponentially, while the dynamics governed by $\mc{H}$ mixes the terms:
\begin{equation}
\begin{aligned}
    e^{t\mathcal{H}}A &= 
    \frac{1}{2}\begin{pmatrix}
    a + d &  b + c \\
    b + c  & a + d 
\end{pmatrix} 
+  \frac{1}{2}\cos(2t) \begin{pmatrix}
    a - d &  b - c \\
   c -  b  & d - a
\end{pmatrix} \\
&+\frac{i}{2}\sin(2t) \begin{pmatrix}
   c -  b &  d - a  \\
   a - d   & b - c
\end{pmatrix}.
\end{aligned}
\end{equation}
We can once again adopt the Trotterization viewpoint. At each infinitesimal time step, the dissipative component $\mathcal{D}$ damps the contributions outside its kernel, which are the off-diagonal elements in this case. Meanwhile, the Hamiltonian component $\mathcal{H}$ induces rotations and mixings among the terms, affecting both those within and outside the kernel of $\mathcal{D}$. This occurs unless $a = d$ and $b = c$. Continuing this iterative process, the system will eventually reach a global equilibrium characterized by a diagonal matrix, where all diagonal terms are equal.

\medskip
\noindent \textit{Theoretical framework and Main Results.---}
We denote by $\sigma$ an equilibrium state of the dynamics $e^{t\mc{L}^\star }$ in the Schr\"{o}dinger picture, and the corresponding Gelfand-Naimark-Segal (GNS) inner product is given by 
\begin{equation} \label{eq:gns_def}
    \average{A, B} = \tr(\sigma A^{\dagger} B).  
\end{equation}
Note that the equilibrium state does not need to be unique. Theorem~1 holds regardless of whether $\sigma$ is full-rank, but to obtain bounds on the mixing time we will need $\sigma$ to be full-rank. This also indicates that if $\sigma$ is full-rank and all the conditions in Theorem~\ref{thm:main} are satisfied, then $\sigma$ must also be the unique equilibrium state, i.e., the quantum Markov semigroup is primitive.
Moreover the norm induced by the GNS inner product is
$    \norm{A} = \average{A, A}^{1/2}.
$
We remark that the discussion below works equally well for other inner products of the form $\average{A,B}_{\alpha} = \tr(\sigma^{\alpha} A^{\dagger}\sigma^{1-\alpha} B)$, although the specific constants in each condition may depend on the inner product. 

In this work we will use three different notions of an adjoint operator: for a super-operator $\mc{M}$, we denote by $\mc{M}^{\star}$ its adjoint under the Hilbert-Schmidt inner product, and by $\mc{M}^*$ its adjoint under the GNS inner product. For an operator $M$, we denote by $M^{\dagger}$ its Hermitian adjoint.

Since the Lindblad equation is linear, the ``difference" between a Hermitian $A$  and the fixed point of $\mc{L}$ starting from $A$, i.e. the fluctuation around the global equilibrium, can be defined as
\[
A - \frac{\langle I, A \rangle}{\langle I,I \rangle}I = A - \tr[\sigma A] I,
\]
which still satisfies the Lindblad equation. Henceforth, all the $A$ we consider will be assumed in the form of such difference.
Note that these include all Hermitian $A$ such that $\tr[\sigma A]=0$.

Let us now present an abstract framework for the convergence of Lindblad equation, largely adapted from hypocoercivity theory for linear kinetic equations~\cite{DolbeaultMouhotSchmeiser2015}.
We will consider the case that $\ker \mc{D}$ is non-trivial (mimicking local equilibria in the kinetic theory), and as a result the dynamics induced by $\mc{D}$ alone is reducible\REV{, meaning that its null space has dimension larger than $1$ or that the fixed point within the null space is not of full rank}. 
We denote by $\mc{P}$ the orthogonal projection onto $\ker \mc{D}$. Because of the inner product structure we have
\begin{equation}
\label{eq:projection_reduces_norm}
    \|(\mc{I}-\mc{P})A\|,\|\mc{P} A\|\leq \|A\|.
\end{equation}
The first condition is that there is a gap separating $0$ (which can be degenerate) from the rest of the eigenvalues of $\mathcal{D}$. %
It ensures that $\mc{D}$ will be able to exponentially damp everything that is not in its kernel. We note that this condition differs from the usual definition of the spectral gap, where the norm on right-hand side needs to be $\norm{A}$. In our formulation, it only needs to be lower bounded by contributions excluding those from $\ker\mathcal{D}$. This accounts for possible degeneracies in the operator $\mathcal{D}$.
\begin{assump}\label{assump:micro}
        The operator $\mc{D}$ \REV{is symmetric, i.e., $\mathcal{D}=\mathcal{D}^*$, and} satisfies 
    $
         - \average{ \mc{D} A, A} \geq \lambda_m \norm{ (\mc{I} - \mc{P}) A}^2, 
    $
    with some positive $\lambda_m$, for all $A$ such that $\tr[\sigma A]=0$.
\end{assump}

We then consider the action of the Hamiltonian on the kernel of $\mc{D}$ and assume the following. \begin{assump}\label{assump:macro}
    The operator $\mc{H}$ is skew-symmetric  and satisfies 
    $\norm{ \mc{H} \mc{P} A}^2 \geq \lambda_M \norm{ \mc{P} A}^2$, with some positive $\lambda_M$, for all $A$ such that $\tr[\sigma A]=0$.
\end{assump}
\REV{This condition implies that the Hamiltonian part will ``stir" the system, just like in the intuitive example, which means that the kernel of $\mc{D}$ will not remain invariant under $\mc{H}$. The constant $\lambda_M$ quantifies this stirring effect.} A straightforward calculation (see \cite[Equation (6)]{thesupplement}) shows that $\mc{H}$ is skew-symmetric if the equilibrium state $\sigma$ commutes with $H$ ($H \sigma = \sigma H$).

We also assume that $\mc{H}$ restricted on $\ker \mc{D}$ vanishes, i.e., 
\begin{assump} \label{assump:C}
    $
        \mc{P} \mc{H} \mc{P} = 0.
    $
\end{assump}
We note that Conditions \ref{assump:macro} and \ref{assump:C} together ensures that $\mc{H}$ will takes outside of $\ker\mc{D}$ any part of $A$ that is orthogonal to $I$ but is inside $\ker\mc{D}$. \REVsec{This condition is typically satisfied by common noise types such as dephasing, bit-flip, and depolarizing noises, but not all Lindbladian dynamics.} See also conclusive remarks for some further remarks on the conditions.

Finally, we make an assumption on the boundedness of $\mc{H} (\mc{I} - \mc{P})$ and $\mc{D}$, which is of course guaranteed for finite dimensional systems, while the bounds will enter into the final estimate. Note that in the $N$-body examples to be discussed later, the constant $C'_M$ in this estimate can be polynomially dependent on the system size $N$ for a $N$-body quantum system, but it does not exhibit exponential dependence.

\begin{assump}
\label{assump:Abound_new}
    For all Hermitian $A$,
    \[
    \|\mc{H}(\mc{I}-\mc{P})A\| + \|\mc{D}A\|\leq C_M'\|(\mc{I}-\mc{P})A\|.
    \]
\end{assump}
We note that a more relaxed condition, as proposed in the supplement~\cite{thesupplement}, still supports our main result. For simplicity, we use \cref{assump:Abound_new} in the main text, which is also easy to verify for our physical examples.

We state our main theorem regarding the convergence rate as follows.
\begin{thm}[Main result]\label{thm:main}
    Under conditions \ref{assump:micro}, \ref{assump:macro}, \ref{assump:C}, and \ref{assump:Abound_new}, there exist positive constants $\lambda$ and $C$, explicitly computable in terms of $\lambda_m$, $\lambda_M$ and $C_M$ such that 
    \begin{equation} \label{eq:thm_main}
        \norm{e^{t ( \mc{H} + \mc{D})} A} \leq C e^{-\lambda t} \norm{A}, \qquad \forall t \geq 0.
    \end{equation}
\end{thm}

With the above we arrive at the conclusion with the parameters given by 
    \begin{equation}
    \label{eq:choice_of_params_C_lam}
    C = \Bigl( \frac{1 + \veps}{1 - \veps}\Bigr)^{1/2}, \quad
    \lambda = \min \left\{\frac{1}{4}\frac{\lambda_m}{1+\veps}, \frac{1}{3} \frac{\veps}{1+\veps} \frac{\lambda_M}{\alpha+ \lambda_M} \right\},
    \end{equation}  
    where $\veps$ are defined in 
   \begin{equation}\label{eq:rate_eps} 
    \veps = \frac{1}{2}\min\left\{\frac{\lambda_m \lambda_M}{(\alpha + \lambda_M) (1+ C_M'/(2\sqrt{\alpha}))^2}, 1 \right\}.
\end{equation}

\noindent \textit{Key proof idea}: To characterize the convergence we construct a twisted norm, which serves as a Lyapunov functional of the system, as 
\begin{equation}
    \mf{L}[A] := \frac{1}{2} \norm{A}^2 - \veps \Re \average{ \mc{A} A, A }, 
\end{equation}
with some $\veps \in (0, 1)$ to be fixed and 
\begin{equation}
    \mc{A} := \bigl(\alpha\mc{I} + (\mc{H} \mc{P})^{\ast}(\mc{H} \mc{P}) \bigr)^{-1} ( \mc{H} \mc{P})^{\ast},
\end{equation}
for some $\alpha>0$. Importantly, this Lyapunov functional can be shown equivalent to $\norm{A}$, and therefore the convergence rate of $\mf{L}[A]$ can be used to estimate the convergence rate of $\|A\|$. See \cite[Section II]{thesupplement} for the proof.

The convergence of observables in the GNS-norm implies convergence of the quantum state in the Schr\"{o}dinger picture, and we can estimate the mixing time as a direct consequence.
\begin{defn}[Mixing time]
\label{defn:mixing_time}
    For a Lindbladian operator $\mc{L}$, we define its $\epsilon$-mixing time to be
    \begin{equation}
        t_{\mathrm{mix}}(\epsilon) = \inf\{t \geq 0:\|e^{s \mc{L}^\star}(\rho)-\sigma\|_1\leq \epsilon,\forall \rho, s\geq t\}.
    \end{equation}
\end{defn}
We have an upper bound for the mixing time (see \cite[Cor 4]{thesupplement} for the derivation):
\begin{equation}
    t_{\mathrm{mix}}(\epsilon)\leq \sup_{\rho}\frac{1}{\lambda}\log(C\|\sigma^{-1}\rho\|/\epsilon)=\frac{1}{\lambda}\log(C\|\sigma^{-1}\|_{\infty}/\epsilon).
\end{equation}

In the limit of small $\lambda_m$, $\lambda_M$, direct calculation shows that the optimal choice for $\alpha$ is $\alpha=\Theta(\lambda_M)$. %
\REV{Here $\Theta(\cdot)$ means there exists constants $0<c_1 <c_2$ such that $c_1\lambda_M\leq \alpha\leq c_2\lambda_M$.}
We have the following mixing time upper bound:
\begin{cor}[Mixing time estimate]
\label{coro:mixing_time_optimized}
    Under conditions \ref{assump:micro}, \ref{assump:macro}, \ref{assump:C}, and \ref{assump:Abound_new}, for $\lambda_m,\lambda_M\leq\Or(1)$, $C_M'\geq\Omega(1)$, if $\sigma$ is full-rank, the mixing time $t_{\mathrm{mix}}(\epsilon)$ defined in \cref{defn:mixing_time} satisfies
    \[
    t_{\mathrm{mix}}(\epsilon)=\Or\left(\frac{C_M'^2}{\lambda_m \lambda_M}\log(\|\sigma^{-1}\|_{\infty}/\epsilon)\right).
    \]
\end{cor}

\noindent \textit{Physical Examples.---}
We now provide a number of many-body physical examples, and apply our framework to provide the mixing time estimates. Note that the $N$-qubit examples can be surprisingly difficult to estimate using other existing frameworks, oftentimes requiring highly non-trivial numerical computation \cite{BarthelZhang2022superoperator}.
We first use a single-qutrit example to show that our framework can deal with the case where the equilibrium state is not the maximally mixed state. 

\medskip
\noindent \textit{Single Qutrit}:
The Hilbert space is spanned by $\{\ket{0},\ket{1},\ket{2}\}$.
We introduce a lowering operator $\sigma^-=\ket{0}\bra{1}$ and a raising operator $\sigma^+=\ket{1}\bra{0}$. We consider the dissipative generator with $V_1=\sqrt{\gamma}\sigma^-$ and $V_2=\sqrt{\gamma}\sigma^+$ (using the notation in \eqref{eq:lindblad_heisenberg}). Then we let the Hamiltonian be $H=\omega(\ket{1}\bra{2}+\ket{2}\bra{1})$. The unique equilibrium state is $\sigma = \frac{1}{2}\ket{0}\bra{0} + \frac{1}{4}\ket{1}\bra{1} + \frac{1}{4}\ket{2}\bra{2}$. 
One can verify all conditions with $\lambda_m=(3/2)\gamma$, $\lambda_M=\omega^2$, and $C_M'=\Or(|\omega|+\gamma)$.
Therefore by \cref{coro:mixing_time_optimized} we can show that the mixing time $t_{\mathrm{mix}}(\epsilon)=\Or((\omega^2+\gamma^2)\omega^{-2}\gamma^{-1}\log(1/\epsilon))$.

\medskip
Next we consider an $N$-qubit quantum system, and let $\mathcal{D}$ be describing dephasing noise, i.e.
\[
\mathcal{D}A = \gamma\sum_i (Z_i A Z_i - A).
\]
Then the null space of $\mathcal{D}$ is spanned by $\{Z^{\otimes \vec{b}}: \vec{b}\in\{0,1\}^N\}$, and is therefore $2^N$ dimensional. Intuitively the stationary states are classical probability distributions over $N$-bit strings. 0 is separated from the rest of the spectrums of $\mathcal{D}$ by a gap of $2\gamma$. In fact, all its eigenvectors can be written down as Pauli matrices. The eigenvalues can be computed through a Hadamard transform of the coefficients. See  \cite[Section II]{ChenZhouSeifJiang2022} for details. \cref{assump:micro} is satisfied with $\lambda_m = 2\gamma$. It is proved in the \cite[Section III]{thesupplement} that \cref{assump:C} holds.

\medskip
\noindent \textit{Transverse Field Ising Model}:
We first study the transverse field Ising model
\begin{equation}
    H = \sum_{i=1}^{N-1}Z_i Z_{i+1} + h \sum_{i=1}^N X_i,
\end{equation}
for $h=\Or(1)$.
We show in \cite[Section III.A.]{thesupplement} that \cref{assump:macro} is satisfied with $\lambda_M = 4h^2$, and $
C_M' = 2((N-1)+Nh)+2N\gamma.
$
Applying \cref{coro:mixing_time_optimized}, we have
\begin{equation}
    \label{eq:mixing_time_TFIM}
    t_{\mathrm{mix}}(\epsilon)= \Or\left(\frac{N^2(1+\gamma)^2}{\gamma h^2}(N+\log(1/\epsilon))\right).
\end{equation}

\noindent \textit{The Heisenberg Model:}
Consider the Heisenberg model
\begin{equation}
\label{eq:heisenberg_model}
    H = -\sum_{i=1}^{N-1} (J_x X_i X_{i+1} + J_y Y_i Y_{i+1} + J_z Z_i Z_{i+1}) + h\sum_{i=1}^N X_i.
\end{equation}
We show in \cite[Section III.B.]{thesupplement} that \cref{assump:macro} is satisfied with $\lambda_M = 4h^2$, and $C_M'=\Or(N(1+\gamma))$, for $h=\Or(1)$. With $\lambda_m=2\gamma$ and through \cref{coro:mixing_time_optimized}, we therefore have
\begin{equation}
    \label{eq:mixing_time_Heisenberg}
    t_{\mathrm{mix}}(\epsilon) = \Or\left(\frac{N^2(1+\gamma)^2}{\gamma h^2}(N+\log(1/\epsilon))\right),
\end{equation}
which is the same as \eqref{eq:mixing_time_TFIM} in the big-$\Or$ notation.

\medskip
\noindent \textit{Quantum Walk Under Dephasing Noise:} 
We can use the above framework to analyze the convergence of the continuous-time quantum walk \cite{FarhiGutmann1998,Childs2009universal,AharonovEtAl1993QuantumRandomWalks,AmbainisEtAl2001OneDimsQuantumWalks} on a graph under dephasing noise, and see that the convergence is governed by the spectral gap of the graph Laplacian, which also governs the convergence of a classical random walk. This setup is somewhat similar to the one studied in \cite{AlagicRussell2005}, but the dissipative part we consider is different.

On a $d$-regular connected graph $G=(V,E)$, we denote by $H$ the adjacency matrix. 
\begin{equation}
    H = \sum_{ij} h_{ij}\ket{i}\bra{j},
\end{equation}
where $h_{ij}=1$ if $(i,j)\in E$, and $h_{ij}=0$ otherwise.
The graph Laplacian is $L=dI-H$, which is a positive semi-definite matrix. The smallest eigenvalue of $L$, which is non-degenerate due to connectedness, is $0$. The second smallest eigenvalue we denote by $\Delta$, is at the same time the gap separating the two smallest eigenvalues. $\Delta$ determines the convergence rate towards the uniform superposition on this graph in a classical random walk.

We now consider the quantum walk described by the time evolution operator $e^{-iHt}$, in the presence of dephasing noise. 
As proved in \cite[Section III.C.]{thesupplement}, we have $\lambda_M=2\Delta$ and $C_M'=\Or(d+\gamma N)$, so that the mixing time estimation is 
\begin{equation}
    t_{\mathrm{mix}}(\epsilon) = \Or\left(\frac{(d+\gamma N)^2}{\gamma\Delta}(N+\log(1/\epsilon))\right).
\end{equation}

\noindent \textit{Discussion.---} 
In this work we used the hypocoercivity framework to analyze the mixing property of the quantum Markov semigroup described by the Lindblad equation. Specifically, we focused on the scenario where the dissipative part has many fixed points, but adding a Hamiltonian part makes the fixed point unique. In this sense the dissipative part alone does not determine the mixing property of the dynamics, thus requiring new tools to probe the combined effect of these two parts. The hypocoercivity framework enabled us to obtain quantitative results about the mixing time, and it does not rely on the detailed balance condition, which is in contrast with most of the previous works on the Lindbladian mixing time \cite{KastoryanoTemme2013,KastoryanoBrandao2016,RouzeFrancaAlhambra2024}. We then applied this framework to quantify the mixing times of various physically relevant models.

The hypocoercivity framework opens up the possibility of analyzing a range of different models that previous techniques cannot tackle. In this work most of the examples we have studied have the dephasing noise as the dissipative part and the maximally mixed state as their fixed point. As a next step we will try to apply this framework to analyze Lindbladian dynamics with more interesting fixed points, such as high-temperature Gibbs states. 
We also note that the current main result \cref{thm:main} does not assume the Hilbert space to be finite-dimensional, and therefore it is of interest to apply this framework to analyze models with infinite-dimensional Hilbert spaces. \REV{Our framework is particularly well-suited for scenarios where the dissipative part is reducible, addressing challenges that previous techniques in the literature cannot effectively resolve. For cases where the dissipative part is irreducible with a unique fixed point, identifying explicit quantitative acceleration of mixing from incorporating a suitable Hamiltonian remains an intriguing open question. Additionally, we note that \cref{assump:C} provides a sufficient but not necessary condition for global exponential convergence to a unique fixed point. Relaxing this condition is an ongoing research direction. This study represents a first step toward estimating mixing times without assuming a detailed balanced Lindbladian and requiring a spectral gap condition for the full Lindbladian. We believe this framework has significant potential, and further relaxation of \cref{assump:C} could lead to more general  results.} 

In hypocoercivity literature, there are other approaches (see e.g., \cite{Villani2009, albritton2019variational, cao2023explicit, bernard2022hypocoercivity}) besides the one that we are adopting that may yield tighter estimate, it is of interest to consider applying these other approaches to obtain more accurate estimates for quantum Markov semigroups.

Another pertinent aspect is the irreducibility property of the Lindbladian equation, for which the algebraic conditions have been proposed \cite{Wolf2012, ZhangBarthel2024, Frigerio1978, SchubertPlastow2023, Yoshida2024}. 
We are able to provide quantitative estimates of the mixing time in this scenario, which goes beyond the proof of irreducibility. 

\section*{Acknowledgements}
The authors thank Thomas Barthel, Bowen Li, Tongyang Li, Lin Lin, Cambyse Rouz\'e for valuable input. This work is supported by National Science Foundation via the grant DMS-2347791 (D.F.) and DMS-2309378 (J.L.), the U.S. Department of Energy, Office of Science, National Quantum Information Science Research Centers, Quantum Systems Accelerator (Y.T.), and the U.S. Department of Energy, Office of Science, Accelerated Research in Quantum Computing Centers, Quantum Utility through Advanced Computational Quantum Algorithms, grant no. DE-SC0025572 (D.F.). 

\bibliography{lindblad}

\newpage
\widetext
\appendix
\section{{Supplementary material for Mixing Time of Open Quantum Systems via Hypocoercivity}}

\setcounter{equation}{0}
\setcounter{figure}{0}
\setcounter{table}{0}
\setcounter{page}{1}

\section{Notations}
We will generally denote quantum states using lower case Greek letters (e.g., $\rho$, $\sigma$), operators using capital letters (e.g., $A$, $B$), super-operators using calligraphic fonts (e.g., $\mc{L}$, $\mc{D}$), and functional on operators using sans-serif fonts (e.g., $\mf{L}$, $\mf{D}$). 
In this work we will use three different notions of an adjoint operator: for a super-operator $\mc{M}$, we denote by $\mc{M}^{\star}$ its adjoint under the Hilbert-Schmidt inner product, and by $\mc{M}^*$ its adjoint under the GNS inner product to be introduced in \eqref{eq:gns_def}. For an operator $M$, we denote by $M^{\dagger}$ its Hermitian adjoint.
\vspace{1em}

Recall that the generator of a quantum Markov semigroup $\mc{T}_t = e^{t \mc{L}}$ in the Heisenberg picture can be written in the Lindblad form:
\begin{equation*}
    \mc{L} A = i [ H, A] + \sum_j V_j^{\dag} [A, V_j] + [V_j^{\dag}, A] V_j.
\end{equation*}
This generator describes the dynamics of an observable in the Heisenberg picture. 
We decompose $\mc{L}$ into a Hamiltonian part and a dissipative part: 
\begin{align*} 
    & \mc{H} A := i[H, A]; \\
    & \mc{D} A := \sum_j V_j^{\dag} [A, V_j] + [V_j^{\dag}, A] V_j.
\end{align*}
We denote the equilibrium state of the dynamics $e^{t\mc{L}^\star }$ in the Schr\"dinger picture as $\sigma$, and the corresponding Gelfand-Naimark-Segal (GNS) inner product is given by 
\begin{equation} \label{eq:gns_def}
    \average{A, B} = \tr(\sigma A^{\dagger} B).  
\end{equation}
Moreover the norm induced by the GNS inner product is
\begin{equation} \label{eq:norm}
    \norm{A} = \average{A, A}^{1/2}.
\end{equation}
We remark that the discussion below works equally well for other inner products of the form $\average{A,B}_{\alpha} = \tr(\sigma^{\alpha} A^{\dagger}\sigma^{1-\alpha} B)$, although the specific constants in each condition may depend on the inner product.

If there are multiple stationary states, one can select any of them. We will focus on the situation where the stationary state is unique. Note that under the conditions outlined below, we can show that the QMS $\mathcal{T}_t$ is in fact primitive, meaning $\dim \ker \mathcal{L} = 1$.

Since the Lindblad equation is linear, the ``difference" between a Hermitian $A$  and the fixed point of $\mc{L}$ starting from $A$, i.e. the fluctuation around the global equilibrium, can be defined as
\[
A - \frac{\langle I, A \rangle}{\langle I,I \rangle}I = A - \tr[\sigma A] I,
\]
which still satisfies the Lindblad equation. Henceforth, all the $A$ we consider will be assumed in the form of such difference.
Note that these include all Hermitian $A$ such that $\tr[\sigma A]=0$.

\section{Proof of the Main Theorems}
Let us now present an abstract framework for the convergence of Lindblad equation, largely adapted from DMS hypocoercivity theory for linear kinetic equations \cite{DolbeaultMouhotSchmeiser2015}.

We will consider the case that $\ker \mc{D}$ is non-trivial (mimicking local equilibria in the kinetic theory), and as a result the dynamics induced by $\mc{D}$ alone is reducible.
We denote by $\mc{P}$ the orthogonal projection onto $\ker \mc{D}$. Because of the inner product structure we have
\begin{equation}
\label{eq:projection_reduces_norm}
    \|(\mc{I}-\mc{P})A\|,\|\mc{P} A\|\leq \|A\|.
\end{equation}
The first condition is that there is a gap separating $0$ (which can be degenerate) from the rest of the eigenvalues of $\mathcal{D}+\mc{D}^*$. It ensures that $\mc{D}$ will be able to exponentially damp everything that is not in its kernel. We note that this condition differs from the usual definition of the spectral gap, where the norm on right-hand side needs to be $\norm{A}$. In our formulation, it only needs to be lower bounded by contributions excluding those from $\ker\mathcal{D}$. This accounts for possible degeneracies in the operator $\mathcal{D}$.
\begin{assump}\label{assump:micro}
        The operator $\mc{D}$ satisfies 
    \begin{equation}
         - \Re \average{ \mc{D} A, A} \geq \lambda_m \norm{ (\mc{I} - \mc{P}) A}^2, 
    \end{equation}
    with some positive $\lambda_m$, for all $A$ such that $\tr[\sigma A]=0$.
\end{assump}

We then consider the action of the Hamiltonian on the kernel of $\mc{D}$ and assume the following. \begin{assump}\label{assump:macro}
    The operator $\mc{H}$ is skew-symmetric  and satisfies 
    \begin{equation}
        \norm{ \mc{H} \mc{P} A}^2 \geq \lambda_M \norm{ \mc{P} A}^2, 
    \end{equation}
    with some positive $\lambda_M$, for all $A$ such that $\tr[\sigma A]=0$.
\end{assump}

The following calculation shows that $\mc{H}$ is skew-symmetric if the equilibrium state $\sigma$ commutes with $H$ ($H \sigma = \sigma H$): 
\begin{equation}
    \begin{aligned}
        \average{\mc{H}A, B} & = \average{i [H, A], B}
    =  \tr \Bigl( \sigma (i HA - i AH)^{\dagger} B \Bigr) \\
    & = \tr \Bigl(\sigma A^{\dagger} (-i HB + i BH) \Bigr) \\
    & = - \tr\Bigl(\sigma A^{\dagger} i [H, B] \Bigr) = - \average{A, \mc{H} B}.
    \end{aligned}
\end{equation}
The existence of $\lambda_M$ needs to be established for specific cases. 

We also assume that $\mc{H}$ restricted on $\ker \mc{D}$ vanishes, i.e., 
\begin{assump} \label{assump:C} The following is true:
    \begin{equation}
        \mc{P} \mc{H} \mc{P} = 0.
    \end{equation}
\end{assump}
We note that Conditions \ref{assump:macro} and \ref{assump:C} together ensures that $\mc{H}$ will takes outside of $\ker\mc{D}$ any part of $A$ that is orthogonal to $I$ but is inside $\ker\mc{D}$. \REVsec{We note that this condition is typically satisfied by common noise types such as dephasing noise, bit-flip noise, and depolarizing noise. However, cases like amplitude damping noise require a different framework. In fact, the challenge in handling amplitude damping noise extends beyond this specific condition and stems from the inherent limitations of $L^2$ Heisenberg picture framework in general. Specifically, the issue arises because the fixed point of amplitude damping noise is not of full rank. This prevents the conversion between the GNS inner product (or any similarly defined inner product) in the Heisenberg picture and the trace distance in the Schr\"odinger picture, making neither the direct gap-based approach nor our approach applicable.
}

\medskip 

To characterize the convergence, we define a twisted norm (which will serve as a Lyapunov function) as 
\begin{equation}
    \mf{L}[A] := \frac{1}{2} \norm{A}^2 - \veps \Re \average{ \mc{A} A, A }, 
\end{equation}
with some $\veps \in (0, 1)$ to be fixed and \REV{the auxiliary superoperator $\mc{A}$ is defined as}
\begin{equation}
    \mc{A} := \bigl(\alpha\mc{I} + (\mc{H} \mc{P})^{\ast}(\mc{H} \mc{P}) \bigr)^{-1} ( \mc{H} \mc{P})^{\ast},
\end{equation}
for some $\alpha>0$.
We collect some properties of $\mc{A}$ in the following lemmas. 
\begin{lem}\label{lem:mcA=mcP_mcA}
    We have 
    \begin{align}
        \mc{A} = \mc{P}\mc{A}  
    \end{align}
\end{lem}
\begin{proof}
    Let us consider $A$ and $B$ satisfying $\mc{A} A = B$, then according to definition
    \begin{equation}\label{eq:mid_eq_HPA}
        \bigl(\mc{HP}\bigr)^{\ast} A = \alpha B + (\mc{H} \mc{P})^{\ast}(\mc{H} \mc{P}) B 
    \end{equation}
    and thus 
    \begin{equation*}
        \alpha B = - \mc{P} \mc{H} A + \mc{P} \mc{H}^2 \mc{P} B \in \ker \mc{D}
    \end{equation*}
    which proves that $\mc{A} = \mc{PA}$. 
\end{proof}

\begin{lem}\label{lem:Abound}
    $\mc{A}$ and $\mc{H}\mc{A}$ are bounded: 
    \begin{equation}
        \norm{\mc{H A} A} \leq \norm{(\mc{I} - \mc{P}) A} \qquad \text{and}, \qquad 
        \norm{\mc{A} A} \leq \frac{1}{2\sqrt{\alpha}} \norm{(\mc{I} - \mc{P}) A}. 
    \end{equation}
\end{lem}
\begin{proof}
    As in the proof of \cref{lem:mcA=mcP_mcA}, consider $A$ and $B$ satisfy $\mc{A} A = B$. Taking the inner product of \cref{eq:mid_eq_HPA} with $B$, one has
    \begin{equation*}
        \average{A, \mc{H} \mc{P} B} = \alpha\average{B,B} + \average{\mc{H}\mc{P}B, \mc{H}\mc{P}B},
    \end{equation*}
    and hence thanks to \cref{assump:C} that $\mc{P}\mc{H}\mc{P}= 0$, we have
    \begin{equation} \label{eq:norm_B+HPB}
        \alpha\norm{B}^2 + \norm{\mc{H}\mc{P}B}^2 = \average{(\mc{I}-\mc{P})A, \mc{H} \mc{P} B}.
    \end{equation}
    To prove the first result in the lemma, we make the following estimates
    \begin{equation*}
      \norm{\mc{H}\mc{P}B}^2 \leq \average{(\mc{I}-\mc{P})A, \mc{H} \mc{P} B} \leq \frac{1}{2}\norm{(\mc{I}-\mc{P})A}^2 + \frac{1}{2} \norm{\mc{H} \mc{P} B}^2,
    \end{equation*}
    which concludes the proof as $\mc{H}\mc{P}B = \mc{H}\mc{P}\mc{A} A =  \mc{H}\mc{A} A$ by \cref{lem:mcA=mcP_mcA}.
    To prove the second result in the lemma, we make another estimate of \cref{eq:norm_B+HPB}, namely,
    \begin{equation*}
        \alpha\norm{B}^2 + \norm{\mc{H}\mc{P}B}^2 \leq \frac{1}{4}\norm{(\mc{I}-\mc{P})A}^2 +  \norm{\mc{H} \mc{P} B}^2. \qedhere
    \end{equation*}

\end{proof}
\medskip 

Finally, we make an assumption on the boundedness of $\mc{A H} (\mc{I} - \mc{P})$ and $\mc{A D}$, which is of course guaranteed for finite dimensional systems, while the bounds will enter into the final estimate. Note that in the $N$-body examples discussed in~\cref{sec:physical_examples}, the constant $C_M$ in this estimate can be polynomially dependent on the system size $N$ for a $N$-body quantum system, but it does not exhibit exponential dependence.

\begin{assump}\label{assump:Abound}
    The super-operators $\mc{A H} (\mc{I} - \mc{P})$ and $\mc{A D}$ are bounded: 
    \begin{equation}
    \label{eq:Abound}
        \norm{\mc{AH} (\mc{I} - \mc{P}) A} + \norm{\mc{AD} A} \leq C_M(\alpha) \norm{(\mc{I} - \mc{P}) A}
    \end{equation}
    for some function $C_M: \RR^+\to \RR^+$.
\end{assump}

We state our main theorem regarding the convergence rate as follows.
\begin{thm}[Main result]\label{thm:main}
    Under conditions \ref{assump:micro}, \ref{assump:macro}, \ref{assump:C}, and \ref{assump:Abound}, there exist positive constants $\lambda$ and $C$, explicitly computable in terms of $\lambda_m$, $\lambda_M$ and $C_M$ such that 
    \begin{equation} \label{eq:thm_main}
        \norm{e^{t ( \mc{H} + \mc{D})} A} \leq C e^{-\lambda t} \norm{A}, \qquad \forall t \geq 0.
    \end{equation}
\end{thm}

\begin{proof}
    We define $A(t)=e^{t(\mc{H}+\mc{D})}A$, and 
    we first calculate the time derivative of $\mf{L}[A]$
    \begin{equation}
        \begin{aligned}
            \frac{\ud}{\ud t} \mf{L}[A] & = \frac{1}{2} \average{(\mc{H} + \mc{D}) A, A} + \frac{1}{2} \average{A, (\mc{H} + \mc{D})A} \\
            & \qquad - \veps\Re \average{\mc{A}(\mc{H} + \mc{D}) A, A} - \veps \Re \average{\mc{A} A, (\mc{H}+\mc{D})A } =: -\mf{D}[A]
        \end{aligned}
    \end{equation}
    where the dissipational funciton $\mf{D}$ is defined as 
    \begin{equation}\label{eq:defD}
        \mf{D}[A] := - \Re\average{\mc{D}A, A}  + \veps \Re \average{\mc{AHP} A, A} + \veps \Re \average{\mc{AH(I - P)} A, A} - \veps \Re \average{\mc{HA}A, A}  + \veps \Re \average{\mc{AD} A, A}.
    \end{equation}
    We note that the term $\veps \average{\mc{D}\mc{A} A, A}$ vanishes since $\mc{A} = \mc{P} \mc{A}$ as proved in \cref{lem:mcA=mcP_mcA} so that $\mc{D}\mc{A} = 0$. 

    By \cref{assump:micro}, 
    \begin{equation*}
        - \Re\average{\mc{D}A, A} \geq \lambda_m \norm{(\mc{I} - \mc{P}) A}^2. 
    \end{equation*}
    For the second term in \eqref{eq:defD}, by \cref{assump:macro}, we have 
    \begin{equation*}
        \begin{aligned}
     \Re \Average{\mc{AHP}A, A} & = \Average{\mc{AHP}A, A}  \\ & =  \Average{ (\alpha\mc{I} + (\mc{HP})^{\ast}(\mc{HP}))^{-1} (\mc{HP})^{\ast} (\mc{HP}) A, A} 
         \geq \frac{ \lambda_M}{\alpha + \lambda_M} \norm{\mc{P} A}^2.
        \end{aligned}
    \end{equation*}
    Note that from the above equation we can see that $\Average{\mc{AHP}A, A}$ is real.
    The remaining terms on the right hand side of \eqref{eq:defD} can be bounded using \cref{assump:Abound}: 
    \begin{equation*}
         \veps \Re\average{\mc{AH(I - P)} A, A} - \veps \Re\average{\mc{HA}A, A}  + \veps \Re\average{\mc{AD} A, A} \geq - \veps (1 + C_M(\alpha)) \norm{(\mc{I} - \mc{P})A} \norm{A},
    \end{equation*}
    where we have used the fact that
    \[  
    \Re\langle\mc{H}\mc{A} A, A\rangle \leq \norm{\mc{H}\mc{A} A}\norm{A} 
    \leq \norm{(\mc{I-P}) A}\norm{A},
    \]
    which follows from Lemma~\ref{lem:Abound}.
    Thus, combined together, we get 
    \begin{equation*}
        \mf{D}[A] \geq \Bigl( \lambda_m - \veps ( 1 + C_M(\alpha)) \Bigl(\frac{1}{2\delta}+\frac{\delta}{2}\Bigr) \Bigr) \norm{(\mc{I} - \mc{P}) A}^2 + \veps \Bigl( \frac{\lambda_M}{\alpha + \lambda_M} - (1 + C_M(\alpha)) \frac{\delta}{2} \Bigr) \norm{\mc{P} A}^2 
    \end{equation*}
    for an arbitrary $\delta > 0$ coming from Cauchy-Schwarz inequality
    \[
    \norm{(\mc{I} - \mc{P})A} \norm{A}\leq \frac{1}{2\delta}\norm{(\mc{I} - \mc{P})A}^2 + \frac{\delta}{2}\norm{A}^2 = \Bigl(\frac{1}{2\delta}+\frac{\delta}{2}\Bigr)\norm{(\mc{I} - \mc{P})A}^2 + \frac{\delta}{2}\norm{\mc{P}A}^2
    \]
    Thus by choosing $\delta$ and then $\veps$ small enough, there exists $\kappa > 0$ such that $\mf{D}[A] \geq \kappa \norm{A}^2$. 

In particular, a set of parameters can be chosen as 
\begin{equation}\label{eq:rate_kappa_eps}
    \kappa = \min \left\{\frac{1}{4}\lambda_m, \frac{1}{3} \veps \frac{\lambda_M}{\alpha+ \lambda_M} \right\}, \quad 
    \veps = \frac{1}{2}\min\left\{\frac{\lambda_m \lambda_M}{(\alpha + \lambda_M) (1+ C_M(\alpha))^2}, 1 \right\}, 
\end{equation}
and 
\[
\delta = \min\Bigl\{\frac{4}{3}\frac{\lambda_M}{(\alpha+\lambda_M)(1+C_M(\alpha))},1\Bigr\},
\]
that does not appear explicitly in the final estimate. Here we take minimum in $\delta$ to ensure that $1/(2\delta)+\delta/2\leq 1/\delta$.

    Finally, notice that by Lemma~\ref{lem:Abound}, $\mf{L}[A]$ is equivalent with $\norm{A}$:
    \begin{equation*}
        \frac{1}{2} (1 - \veps) \norm{A}^2 \leq \mf{L}[A] \leq \frac{1}{2} (1 + \veps) \norm{A}^2.
    \end{equation*}
    Therefore
    \[
    \frac{\ud}{\ud t}\mf{L}[A] = -\mf{D}[A]\leq -\kappa\|A\|^2\leq-\frac{2\kappa}{1+\veps}\mf{L}[A].
    \]
    With the above we arrive at the conclusion with the parameters given by 
    \begin{equation}
    \label{eq:choice_of_params_C_lam}
    C = \Bigl( \frac{1 + \veps}{1 - \veps}\Bigr)^{1/2}, \quad
    \lambda = \kappa / (1 + \veps) = \min \left\{\frac{1}{4}\frac{\lambda_m}{1+\veps}, \frac{1}{3} \frac{\veps}{1+\veps} \frac{\lambda_M}{\alpha+ \lambda_M} \right\},
    \end{equation}  
    where $\veps$ are defined in \cref{eq:rate_kappa_eps}.
\end{proof}

We now show that convergence of observables in the GNS-norm implies convergence of the quantum state.
\begin{cor}[Convergence in the Schr\"odinger picture]
\label{coro:state_convergence}
    Under conditions \ref{assump:micro}, \ref{assump:macro}, \ref{assump:C}, and \ref{assump:Abound}, if $\sigma$ is full-rank, for any quantum state $\rho$ (positive semi-definite with $\tr\rho=1$), we have
    \begin{equation}
    \label{eq:convergence_Schrodinger_picture}
    \|e^{t\mc{L}^\star}(\rho)-\sigma\|_1\leq C\|\sigma^{-1}\rho\| e^{-\lambda t},
\end{equation}
for any $t\geq 0$, and with the same $C$ and $\lambda$ as in Theorem~\ref{thm:main}.
\end{cor}

\begin{proof}
At time $t$, the expectation value of $A$ starting from a state $\rho$ at time $0$ is
\[
\tr[\rho A(t)] = \tr[\sigma \sigma^{-1}\rho A(t)] = \average{\sigma^{-1}\rho, A(t)}.
\]
As $A(t)\to 0$ we want to show how fast $\average{\sigma^{-1}\rho, A(t)}$ approaches 0. This can be done via the Cauchy-Schwarz inequality.
\[
|\average{\sigma^{-1}\rho, A(t)}|\leq \|\sigma^{-1}\rho\|\|A(t)\|\leq C\|\sigma^{-1}\rho\| e^{-\lambda t}\|A\|,
\]
for any Hermitian $A$ such that $\tr[\sigma A]=0$. For a generic Hermitian $A$, we can replace $A$ with $A-\tr[\sigma A]I$ in the inequality above to get
\[
|\tr[\rho A(t)] - \tr[\sigma A]|\leq C\|\sigma^{-1}\rho\| e^{-\lambda t}\|A-\tr[\sigma A]I\|.
\]
We can switch to the Schr\"{o}dinger picture, and observe that $\tr[\rho A(t)]=\tr[e^{t\mc{L}^\star}(\rho)A]$. Therefore
\[
|\tr[e^{t\mc{L}^\star}(\rho)A] - \tr[\sigma A]|\leq C\|\sigma^{-1}\rho\| e^{-\lambda t}\|A-\tr[\sigma A]I\|\leq C\|\sigma^{-1}\rho\| e^{-\lambda t}\|A\|_{\infty},
\]
where we have used $\|A-\tr[\sigma A]I\|\leq \|A\|\leq \|A\|_{\infty}$.
Note that this is true for every Hermitian $A$, and therefore we have \eqref{eq:convergence_Schrodinger_picture}.
\end{proof}

From the convergence in the Schr\"{o}dinger picture we can estimate the mixing time.
\begin{defn}[Mixing time]
\label{defn:mixing_time}
    For a Lindbladian operator $\mc{L}$, we define its $\epsilon$-mixing time to be
    \begin{equation}
        t_{\mathrm{mix}}(\epsilon) = \inf\{t \geq 0:\|e^{s \mc{L}^\star}(\rho)-\sigma\|_1\leq \epsilon,\forall \rho, s\geq t\}.
    \end{equation}
\end{defn}
\cref{coro:state_convergence} enables us to provide an upper bound for the mixing time, which is
\begin{equation}
    t_{\mathrm{mix}}(\epsilon)\leq \sup_{\rho}\frac{1}{\lambda}\log(C\|\sigma^{-1}\rho\|/\epsilon)=\frac{1}{\lambda}\log(C\|\sigma^{-1}\|_{\infty}/\epsilon).
\end{equation}
We note that in the above expression, the mixing time is primarily determined by $\lambda$ from \cref{thm:main}. In the theorem, there is a parameter $\alpha$ that we have not chosen, and we will now optimize $\alpha$ to obtain a tighter bound for the mixing time in a more compact form. To this end we will make explicit the dependence on $\alpha$ in the constant $C_M(\alpha)$ introduced in \cref{assump:Abound} by introducing a new assumption:
\begin{assump}
\label{assump:Abound_new}
    For all Hermitian $A$,
    \[
    \|\mc{H}(\mc{I}-\mc{P})A\| + \|\mc{D}A\|\leq C_M'\|(\mc{I}-\mc{P})A\|.
    \]
\end{assump}
This assumption, combined with \cref{lem:Abound}, then implies that
\[
\|\mc{A}\mc{H}(\mc{I}-\mc{P})A\| + \|\mc{A}\mc{D}A\|\leq \frac{1}{2\sqrt{\alpha}}C_M'\|(\mc{I}-\mc{P})A\|.
\]
Therefore we can choose $C_M(\alpha)=\frac{1}{2\sqrt{\alpha}}C_M'$. 

In the limit of small $\lambda_m$, $\lambda_M$, direct calculation shows that the optimal choice for $\alpha$ is $\alpha=\Theta(\lambda_M)$. We then have from \eqref{eq:choice_of_params_C_lam}
\[
C=\Theta(1),\quad \lambda= \Theta\left(\frac{\lambda_m \lambda_M}{C_M'^2}\right).
\]
Therefore we have the following mixing time upper bound:
\begin{cor}
\label{coro:mixing_time_optimized}
    Under conditions \ref{assump:micro}, \ref{assump:macro}, \ref{assump:C}, and \ref{assump:Abound_new}, for $\lambda_m,\lambda_M=\Or(1)$, $C_M'=\Omega(1)$, if $\sigma$ is full-rank, the mixing time $t_{\mathrm{mix}}(\epsilon)$ defined in \cref{defn:mixing_time} satisfies
    \[
    t_{\mathrm{mix}}(\epsilon)=\Or\left(\frac{C_M'^2}{\lambda_m \lambda_M}\log(\|\sigma^{-1}\|_{\infty}/\epsilon)\right).
    \]
\end{cor}

\section{Proof of Mixing Time for Physical Examples}
\label{sec:physical_examples}

We can consider an $N$-qubit quantum system, and let $\mathcal{D}$ be describing dephasing noise, i.e.
\[
\mathcal{D}[A] = \gamma\sum_i (Z_i A Z_i - A).
\]
Then the null space of $\mathcal{D}$ is spanned by $\{Z^{\otimes \vec{b}}: \vec{b}\in\{0,1\}^N\}$, and is therefore $2^N$ dimensional. Intuitively the stationary states are classical probability distributions over $N$-bit strings. 0 is separated from the rest of the spectrums of $\mathcal{D}$ by a gap of $2\gamma$. In fact, all its eigenvectors can be written down as Pauli matrices. The eigenvalues can be computed through a Hadamard transform of the coefficients. See  \cite[Section II]{ChenZhouSeifJiang2022} for details. \cref{assump:micro} is satisfied with $\lambda_m = 2\gamma$.

We next prove that \cref{assump:C} holds, i.e., $\mc{P}\mc{H}\mc{P}=0$, with this $\mc{D}$ for all Hamiltonians $H$, if the equilibrium state is $\sigma=I/2^N$. 
We only need to show that for any pair of Z-strings $Q$ and $Q'$, 
\begin{equation}
    \label{eq:verify_condition_C}
    \langle Q,\mc{H}Q'\rangle = \langle Q,[H,Q']\rangle=\frac{1}{2^N}\tr[Q[H,Q']]=0,
\end{equation}
because Z-strings form a basis of $\Im\mc{P}$.
Note that a Hamiltonian can be written as a linear combination of Pauli matrices
\[
H = \sum_j P_j.
\]
Therefore we have
\[
\frac{1}{2^N}\tr[Q[H,Q']] = \frac{1}{2^N}\sum_j\tr[Q[P_j,Q']].
\]
Because $Q'$ is a Z-string, then $[P_j,Q']$ is therefore either 0 or a Pauli string that contains a non-Z component, and is therefore orthogonal to the Z-string $Q$. In either case we have
\[
\sum_j\tr[Q[P_j,Q']] = 0.
\]
Therefore we have proved that $\langle Q,\mc{H}Q'\rangle=0$ for any Z-strings $Q$ and $Q'$, and consequently \cref{assump:C} holds.

\subsection{Transverse Field Ising Model} \label{sec:ex_tfim}
We first study the transverse field Ising model
\begin{equation}
    H = \sum_{i=1}^{N-1}Z_i Z_{i+1} + h \sum_{i=1}^N X_i.
\end{equation}
We would like to check the conditions proposed in the theoretical framework, and show that $I/2^N$ is the unique global equilibrium of $L$ and the convergence rate towards this equilibrium is constant.

Note that we have already checked \cref{assump:micro} and \cref{assump:C} at the beginning of \cref{sec:physical_examples}.
Next, we show that \cref{assump:macro} is satisfied with $\lambda_M = 4h^2$. 

Hereafter, $\|\cdot \|_{\mathrm{NHS}}$ denotes the Hilbert-Schmidt norm normalized with $2^{N/2}$ so that each Pauli matrix has norm $1$. Note that in our case, this normalized Hilbert-Schmidt norm coincides with the norm induced by the GNS inner product \cref{eq:gns_def}.

We start by considering a matrix $A \in \ker(\mathcal{D})$ with $\tr{A} = 0$ and $\norm{A}_\mathrm{NHS} = 1$ and hence we can write $A = \sum_j a_j P_j$, where $\sum_j |a_j|^2 = 1$ and $P_j$'s are Z-strings. Moreover, $H$ can be written as $H = \sum_k h_k Q_k$, where $h_k$ is either $1$ or $h$, and $Q_k$ is in the form of either $Z_i Z_{i+1}$ or $X_i$. 

One helpful observation is that the commutators $\left\{ [Q_k, P_j] \right\}_{k, j}$ are orthogonal to each other under the Hilbert-Schmidt inner product, namely,
\begin{equation}
\label{eq:orthogonality_commutator}
    \tr \bigl[[Q_k, P_j]^{\dagger}[Q_{k'}, P_{j'}]\bigr] = 0,
\end{equation}
for $(k, j) \neq (k',j')$.
To see this, we first note that all $Z_i Z_{i+1}$'s commute with $P_j$'s, and only when $Q_k$ are of the form $X_i$, the commutator $[Q_k, P_j]$ may be non-zero.
Therefore, it is sufficient to consider $Q_k$ that are in the form of $X_i$. To make
\begin{equation}
     \tr\left( [Q_k, P_j]^{\dagger}[Q_{k'}, P_{j'}] \right) = 
     - 4\tr\left(Q_k P_j P_{j'} Q_{k'} \right) = - 4\tr\left( Q_{k'} Q_k P_j P_{j'}  \right) \neq 0
\end{equation}
holds, the only possibility is when the two Pauli strings $Q_{k'}Q_k = P_jP_{j'} = I$ are the same, which implies $k' = k$ and $j' = j$.

Therefore, we have
\[
\|[H,A]\|_{\mathrm{NHS}}^2 = \sum_{kj} |h_k a_j|^2 \|[Q_k, P_j]\|_{\mathrm{NHS}}^2.
\]
As $P_j \neq I$, there exists at least one $Z$ in $P_j$, and hence $P_j$ anti-commutes with at least one $X_i$ term. We have
\[
\sum_{k} |h_k|^2 \|[Q_k, P_j]\|_{\mathrm{NHS}}^2 \geq 4 h^2,
\]
which implies 
\[
\norm{\mathcal{H} A}^2 = \|[H,A]\|_{\mathrm{NHS}}^2 \geq 4h^2, 
\]
for $A \in \ker(\mathcal{D})$ with $trA=0$ and $\norm{A} =\norm{A}_\mathrm{NHS} = 1$.
We can conclude that \cref{assump:macro} is satisfied with $\lambda_M = 4h^2$.

We will now compute the parameter $C_M'$ in \cref{assump:Abound_new}. 
Note that this task is much easier than the general case because $\sigma=I/2^N$ and we can utilize the nice properties of the Hilbert-Schmidt norm. In particular, we have
\[
\|\mc{H}(\mc{I}-\mc{P})A\|\leq 2\|H\|_{\infty}\|(\mc{I}-\mc{P})A\|\leq 2(\|H\|_{\infty}\leq (N-1)+Nh)\|(\mc{I}-\mc{P})A\|,
\]
where the first inequality is only true for $\|\cdot\|=\|\cdot\|_{\mathrm{NHS}}$. We have also used the fact that $\|H\|_{\infty}\leq (N-1)+Nh$. 
For the dissipative part, we have
\[
\|\mc{D}A\|=\|\mc{D}(\mc{I}-\mc{P})A\|\leq \gamma\sum_i (\|Z_i((\mc{I}-\mc{P})A)Z_i\|+\|(\mc{I}-\mc{P})A\|)\leq 2N\gamma\|(\mc{I}-\mc{P})A\|.
\]
We can therefore choose
\[
C_M' = 2((N-1)+Nh)+2N\gamma.
\]
Combining the above with $\lambda_m=2\gamma$, $\lambda_M=4h^2$, and $\|\sigma^{-1}\|_{\infty}=2^N$, and assuming $h=\Or(1)$, we have from \cref{coro:mixing_time_optimized} that
\begin{equation}
    \label{eq:mixing_time_TFIM}
    t_{\mathrm{mix}}(\epsilon) = \Or\left(\frac{N^2(1+\gamma)^2}{\gamma h^2}\log(2^N/\epsilon)\right) = \Or\left(\frac{N^2(1+\gamma)^2}{\gamma h^2}(N+\log(1/\epsilon))\right).
\end{equation}

\subsection{The Heisenberg Model}

We consider 
\begin{equation}
\label{eq:heisenberg_model}
    H = -\sum_{i=1}^{N-1} (J_x X_i X_{i+1} + J_y Y_i Y_{i+1} + J_z Z_i Z_{i+1}) + h\sum_{i=1}^N X_i.
\end{equation}

Because we have verified Conditions \ref{assump:micro} and \ref{assump:C} at the beginning of Section~\ref{sec:physical_examples}, we only need to focus on Conditions \ref{assump:macro} and \ref{assump:Abound_new}.
We now check \cref{assump:macro}.
In this model, we note that \eqref{eq:orthogonality_commutator} is no longer correct: one can check that 
\[
[X_1 X_2, Z_1] = -[Y_1 Y_2,Z_2].
\]
We will see, however, that the $h\sum_{i=1}^N X_i$ part alone is still sufficient for obtaining an upper bound of the mixing time.

We still write $H=\sum_k Q_k$, and $A=\sum_j\alpha_j P_j$, with $\sum_j|\alpha_j|^2=1$. Then we observe that
\begin{equation}
    \label{eq:limited_orthogonality_commutator}
    \tr[[X_i,P_j][Q_k,P_{j'}]]=0
\end{equation}
if $j\neq j'$ or $Q_k\neq X_i$. To see this, we note that if $\tr[[X_i,P_j][Q_k,P_{j'}]]\neq 0$ then $[X_i,P_j]\neq 0$, $[Q_k,P_{j'}]\neq 0$, and $X_i P_j=\pm Q_k P_{j'}$. The last condition can be equivalently written as  $Q_k X_i =\pm P_{j'} P_j$. This means that $Q_k X_i$ is a $Z$-string. By examining each term of the Hamiltonian we notice that this is only possible with $Q_k=X_i$. This then implies that $j=j'$.

We then write
\[
[H,A] = \sum_j \sum_{k:Q_k\notin \{X_i\}} h_k\alpha_j [Q_k,P_j] + h\sum_j \sum_i \alpha_j [X_i,P_j].
\]
Because of \eqref{eq:limited_orthogonality_commutator}, the two parts on the right-hand side are orthogonal to each other under the Hilbert-Schmidt inner product. Therefore
\[
\|[H,A]\|_{\mathrm{NHS}} \geq h\|\sum_j \sum_i \alpha_j [X_i,P_j]\|_{\mathrm{NHS}}.
\]
Again using \eqref{eq:limited_orthogonality_commutator},
\[
\|\sum_j \sum_i \alpha_j [X_i,P_j]\|_{\mathrm{NHS}}^2 = \sum_j\sum_i |\alpha_j|^2\|[X_i,P_j]\|_{\mathrm{NHS}}^2.
\]
Because for each $j$, there exists at least one $i$ such that $X_i$ and $P_j$ anti-commute, thus ensuring $\|[X_i,P_j]\|_{\mathrm{NHS}}=2$, we have
\[
\sum_j\sum_i |\alpha_j|^2\|[X_i,P_j]\|_{\mathrm{NHS}}^2 \geq 4\sum_j |\alpha_j|^2\geq 4.
\]
Therefore, \cref{assump:macro} is satisfied with $\lambda_M = 4h^2$.

For \cref{assump:Abound_new}, the same analysis as in the previous section yields that $C_M'=\Or(N(1+\gamma))$, for $h=\Or(1)$. With $\lambda_m=2\gamma$ and through \cref{coro:mixing_time_optimized}, we therefore have
\begin{equation}
    \label{eq:mixing_time_Heisenberg}
    t_{\mathrm{mix}}(\epsilon) = \Or\left(\frac{N^2(1+\gamma)^2}{\gamma h^2}(N+\log(1/\epsilon))\right),
\end{equation}
which is the same as \eqref{eq:mixing_time_TFIM} in the big-$\Or$ notation.

\subsection{Quantum Walk Under Dephasing Noise}
\label{sec:quantum_walk_dephasing}

We can use the above framework to analyze the convergence of the quantum walk on a graph under dephasing noise, and see that the convergence is governed by the spectral gap of the graph Laplacian, which also governs the convergence of a classical random walk.

On a $d$-regular connected graph $G=(V,E)$, we denote by $H$ the adjacency matrix. 
\begin{equation}
    H = \sum_{ij} h_{ij}\ket{i}\bra{j},
\end{equation}
where $h_{ij}=1$ if $(i,j)\in E$, and $h_{ij}=0$ otherwise.
The graph Laplacian is $L=dI-H$, which is a positive semi-definite matrix. The smallest eigenvalue of $L$, which is non-degenerate due to connectedness, is $0$. The second smallest eigenvalue we denote by $\Delta$, is at the same time the gap separating the two smallest eigenvalues. $\Delta$ determines the convergence rate towards the uniform superposition on this graph in a classical random walk.

We now consider the quantum walk described by the time evolution operator $e^{-iHt}$, in the presence of dephasing noise. By the same analysis as in previous examples, the set of fixed points of the dephasing channel is all the $Z$-strings. Therefore we only need to consider
\[
A = \sum_{i=0}^{2^N-1} a_i \ket{i}\bra{i},
\]
such that $\sum_i a_i=0$, $\frac{1}{2^N}\sum_i a_i^2=1$.
We then compute that
\[
[H,A] = \sum_{ij} h_{ij}(a_j-a_i)\ket{i}\bra{j}.
\]
Therefore
\[
\|[H,A]\|_{\mathrm{NHS}}^2 = \sum_{ij} h_{ij}^2(a_j-a_i)^2 = \frac{2\vec{a}^\top L \vec{a}}{2^N},
\]
where $\vec{a}=(a_1,a_2,\cdots,a_{2^N})$, and we have used the fact that $h_{ij}=h_{ji}$, and $\sum_{j}h_{ij}=d$. Because $\sum_i a_i=0$, $\vec{a}$ is thus orthogonal to the lowest eigenstate of $L$, and consequently we have $\vec{a}^\top L \vec{a}\geq 2^N\Delta$. Therefore
\[
\|[H,A]\|_{\mathrm{HS}}^2 \geq 2\Delta.
\]
We can therefore choose $\lambda_M=2\Delta$. Using the fact that $\|H\|_\infty=d$, we have $C_M'=\Or(d+\gamma N)$ in \cref{assump:Abound_new}. Therefore, through \cref{coro:mixing_time_optimized}, the mixing time is
\begin{equation}
    t_{\mathrm{mix}}(\epsilon) = \Or\left(\frac{(d+\gamma N)^2}{\gamma\Delta}(N+\log(1/\epsilon))\right).
\end{equation}

\REVsec{\section{Numerical Results}}

\begin{figure}
    \centering
    \includegraphics[width=0.6\linewidth]{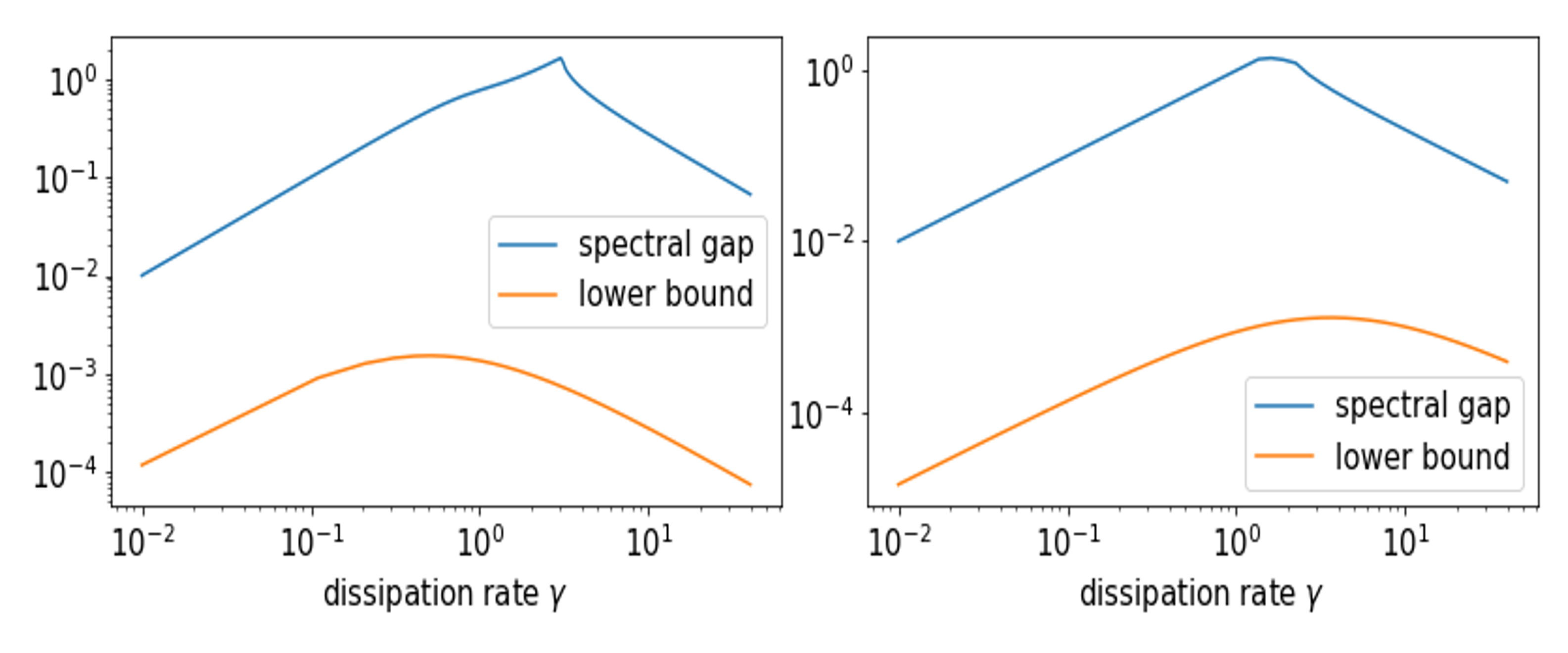}
    \caption{\REV{The spectral gap (computed numerically) and the lower bound $\lambda$ (computed using \eqref{eq:choice_of_params_C_lam}) as functions of the dissipation strength $\gamma$ for the single-qutrit example (left) and the Heisenberg model (right) under dephasing noise as discussed below. The parameters are chosen to be $\omega=J_x=J_y=J_z=h=1$, and the Heisenberg model consists of $N=4$ sites.}}
    \label{fig:gap_vs_lower_bound}
\end{figure}

\REVsec{In Figure~\ref{fig:gap_vs_lower_bound} we compare $\lambda$ computed through \eqref{eq:choice_of_params_C_lam}, which lower bounds the spectral gap of the Lindbladian, with the exact spectral gap, defined to be the absolute value of the largest real part of the non-zero eigenvalues of the Lindbladian, which governs the convergence rate asymptotically at large times. The numerics indicate that the lower bounds thus computed are not tight, but they nonetheless reflect the general trend of how the gap changes with the dissipation strength: with a large $\gamma$ every $Z$-string becomes almost a fixed point, making it difficult to escape from it, thus resulting in a diminishing spectral gap; with small $\gamma$ the dissipative part is too weak, which also results in a diminishing spectral gap.}

\end{document}